# Inter-Comparison and Validation of Geant4 Photon Interaction Models

M. Augelli, M. Begalli, M. G. Pia, P. P. Queiroz, L. Quintieri, D. Souza-Santos, M..Sudhakar, P. Saracco, G. Weidenspointner, A. Zoglauer

***Abstract*–A R&D project, named Nano5, has been recently launched to study an architectural design in view of addressing new experimental issues related to particle transport in high energy physics and other related physics disciplines with Geant4. In this frame, the first step has involved the redesign of the photon interaction models currently available in Geant4; this task has motivated a thorough investigation of the physics and computational features of these models, whose first results are presented here.**

## I. Introduction

A project is in progress concerning the systematic evaluation of the simulation models of photon interactions pertinent to the Geant4 toolkit [1][2]. It takes place in the framework of the Nano5 [3][4] R&D project: in that context, software design techniques are explored [5] to evaluate their capability of supporting new concepts in particle transport schemes, like the mutability of physics-related objects.

The investigation of design techniques, motivated by new functional requirements to be introduced in the simulation, provides the ground for estimating quantitatively the effects of the design itself on the software test process, as well as the physics and computational performance of the available Geant4 models. Some preliminary results are reported here.

## II. Prototype Software Design

A pilot project currently in progress evaluates the adoption of generic programming techniques in Geant4 electromagnetic physics. The first development cycle in this context has investigated the use of policy-based class design in the domain of photon interactions.

The prototype design has adopted a minimalist approach: a generic process acts as a host class, which is deprived of any intrinsic physics functionality. Physics behavior is acquired through policy classes, respectively responsible for cross section and final state generation. In this way, the physics process is independent from the models that determines the cross-section and final state generation.

The main characteristics and advantages of this approach are:

- a policy defines a class or a class template interface
- policy host classes are parameterized classes
- policies are not required to inherit from a base class
- the code is bound at compilation time: since no virtual methods are needed, faster execution is expected.

The guiding concepts are:

- flexible configuration of processes at granular level, which is required to distinguish stable and mutable state and behavior
- performance optimization for computationally intensive use cases
- transparency of physics
- effortless Verification & Validation

Figure 1 shows an example of the prototype design of photon interactions, concerning Compton scattering.

Cross Section and Final State are estimated by two specialized concrete classes. A host process class is parameterized over these two policy classes. Cross section and final state policies are orthogonal.

## III. Software Verification and Validation

A side product of the adoption of generic programming techniques in the design is the improved transparency of physics models, that is exposed at a fine-grained level. This feature greatly facilitates the test of the code at microscopic level and the flexible configuration of physics processes in multiple combinations of modeling features.

The testing of basic physics components (atomic cross sections or features of the final state models) is greatly facilitated with respect to the current Geant4. This is due to the fact that these computations are performed by low level objects like policy classes; therefore the final state and cross section implementations can be verified and validated independently, in a simple unit testing environment, whereas in the current design scheme a full scale Geant4 based

Manuscript received November 13, 2009. This work was supported in part by the Conselho Nacional de Desenvolvimento Científico e Tecnológico – CNPq, Brazil.

Maria Grazia Pia, P. Saracco and M..Sudhakar are with INFN, Sezione Genova, Italy (e-mail: MariaGrazia.Pia@ge.infn.it, Paolo.Saracco@ge.infn.it, Manju.Sudhakar@ge.infn.it).

Mauro Augelli is with Centre National d'Études Spatiales (CNES), France (e-mail: mauroaugelli@mac.com).

Lina Quintieri is with INFN-LNF Frascati, Italy (e-mail: Lina.Quintieri@lnf.infn.it).

Georg Weidenspointner is with MPE-MPI, Munich, Germany (e-mail: Georg.Weidenspointner@hll.mpg.de).

Andreas Zoglauer is with University of California, Berkeley.(e-mail: zog@ssl.berkeley.edu)

Pedro P. Queiroz Filho and Denison Souza-Santos are with the Institute for Radiation Protection and Dosimetry – IRD/CNEN, Rio de Janeiro, Brazil (e-mails: queiroz@ird.gov.br and santosd@ird.gov.br).

Márcia Begalli is with the State University of Rio de Janeiro – UERJ, Brazil (e-mail: marcia.begalli@cern.ch).

application is necessary to study even low-level physics entities.

The comparison of the various Geant4 implemented models (Standard, Penelope, Livermore) and their physics accuracy against established reference databases, like the NIST data collection, and experimental data available in literature is in progress. A first set of results is discussed here.

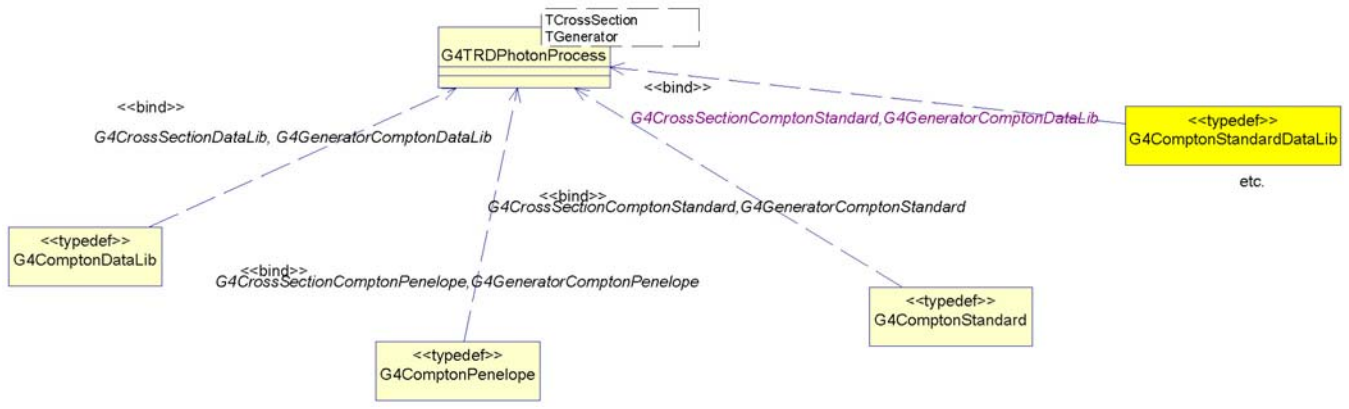

Figure 1: UML diagram of the prototype photon interactions design.

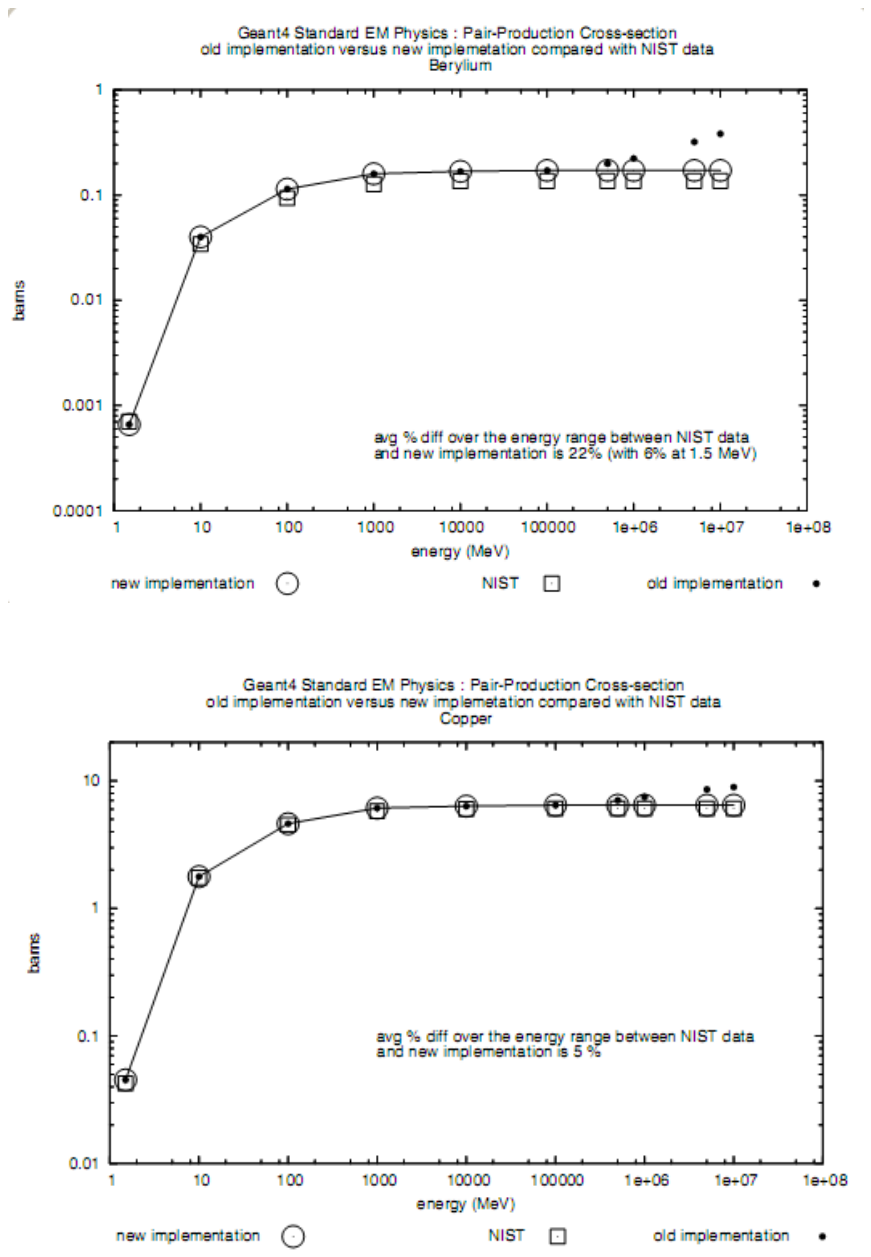

Figure 3: Photon conversion cross-section validation with respect NIST reference data for Be and Cu.

The plots in Fig. 3 refer to the case of photon conversion cross-section comparison with respect to NIST reference data for two cases: Be and Cu.

Discrepancies between the implementation in Geant4 9.2 standard electromagnetic package and the User Documentation of the cross-section model for photon conversion have been identified and reported to the maintainers of the original Geant4 implementations. The observed model behaviour is shown in Fig. 4. According to the Geant4 Physics Manual, above 100 GeV the cross-section for photon conversion, based on the Bethe-Heitler model, should be constant.

The agility of Nano5 electromagnetic physics design allowed the implementation of a version of the cross-section computation consistent with the specifications of Geant4 Physics Reference Manual, as an alternative to the existing one. The provision of an alternative option and its interchangeability in a user application are straightforward in the proposed software design environment.

Other implementations of this cross section based on alternative models documented in literature are in progress; for instance, according to [1] the cross section above 1 TeV is expected to fall as a function of energy. Implementations by Geant4 standard electromagnetic group have been announced in Geant4 development plans.

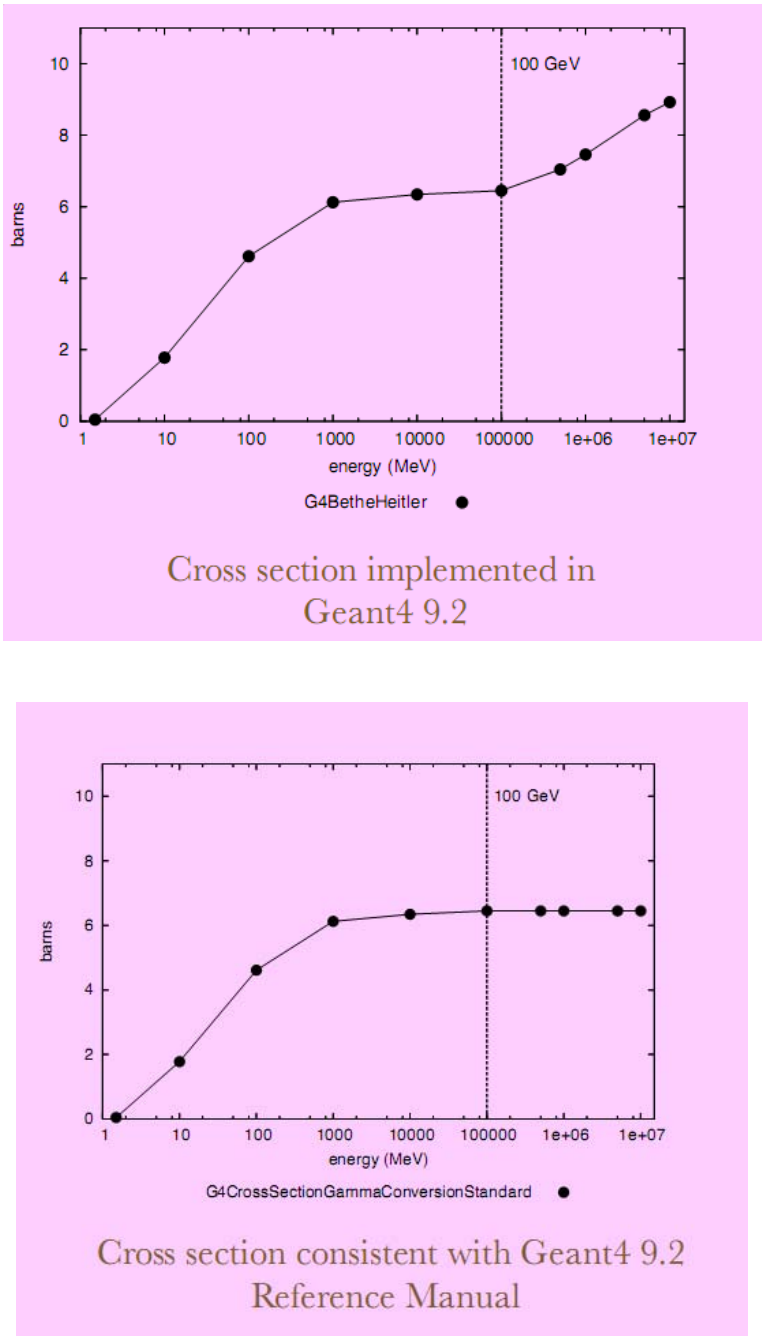

Figure 4: Differences between the implementation in Geant4 9.2 standard electromagnetic package and the User Documentation of the cross-section calculation for photon conversion

All photon interactions are implemented in Geant4 in at least three modeling variants, which are identified as Standard, library-based and Penelope-like. The minimalistic approach of the policy based class design greatly facilitates the relative comparison of Geant4 physics models in quantitative terms: their physics features and computational performance.

The example given in Fig. 5, concerning Compton scattering, shows that the Library-based and Penelope-like cross sections exhibit similar characteristics over the 1 keV – 100 GeV energy range; the Standard cross section model shows larger differences with respect to the Library-based one.

Similarly, the features of the various final state modelling options can be easily compared. Fig. 7 shows the distribution of the ratio between the initial and final photon energy resulting from the final state generators of the Library, Penelope and Standard flavors of Compton scattering, for two different target elements and incident photon energies.

The accuracy with respect to the NIST reference data has been estimated quantitatively by goodness-of-fit testing. An example is given in Table 1, which reports the p-value of the $\chi^2$ test concerning Compton scattering cross sections over the 1 keV – 100 GeV energy range; the Library-based cross section model appear to provide better accuracy with respect to the NIST Physics Reference Data.

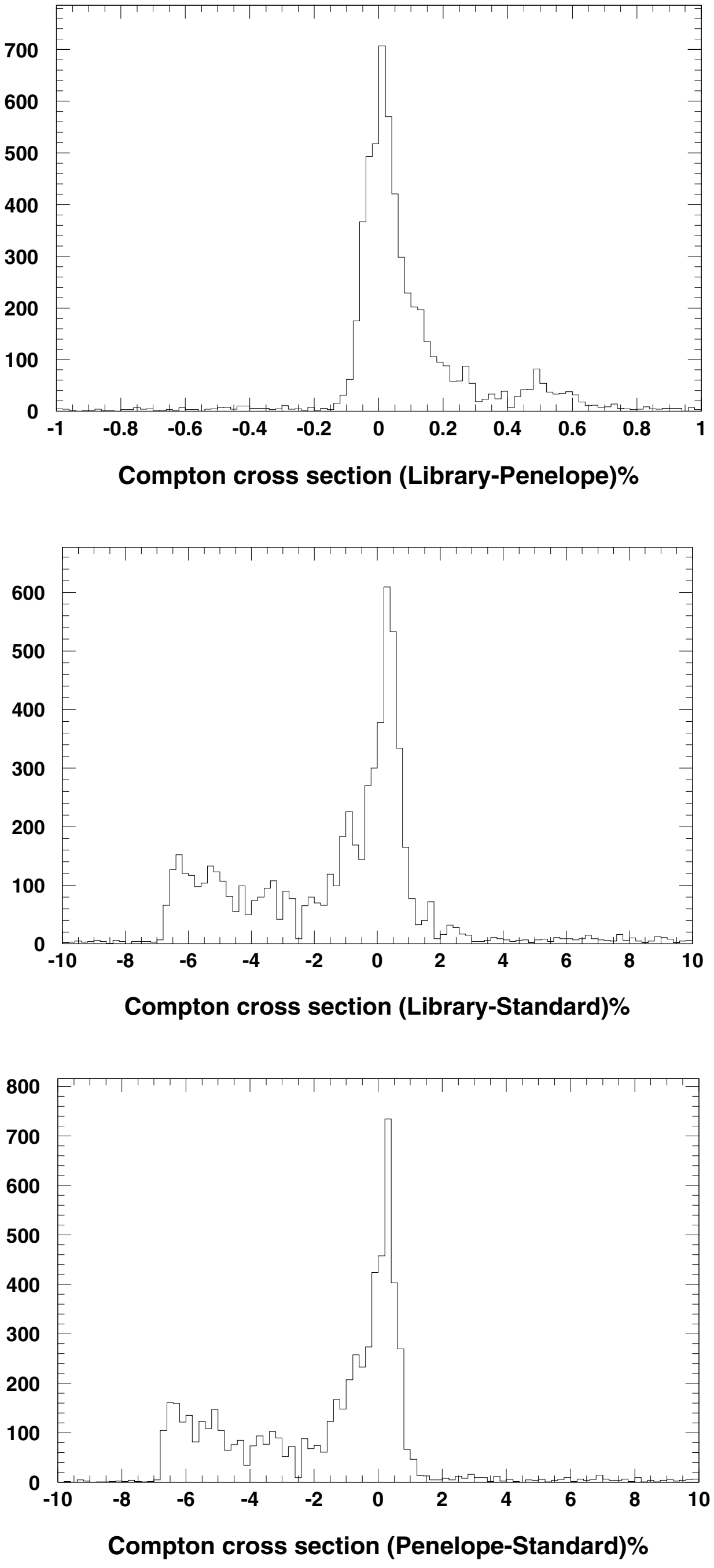

Figure 5: Behavior of the three photon interaction models in Geant4 with respect to the NIST reference data.

These results provide objective guidance to experimentalists in the choice of Geant4 physics features to be selected in their simulation applications (in users' PhysicsLists). They also provide indications to Geant4 managers for possible optimization of Geant4 maintenance.

Further tests are in progress for each photon interaction type regarding the validation of the various models against experimental data.

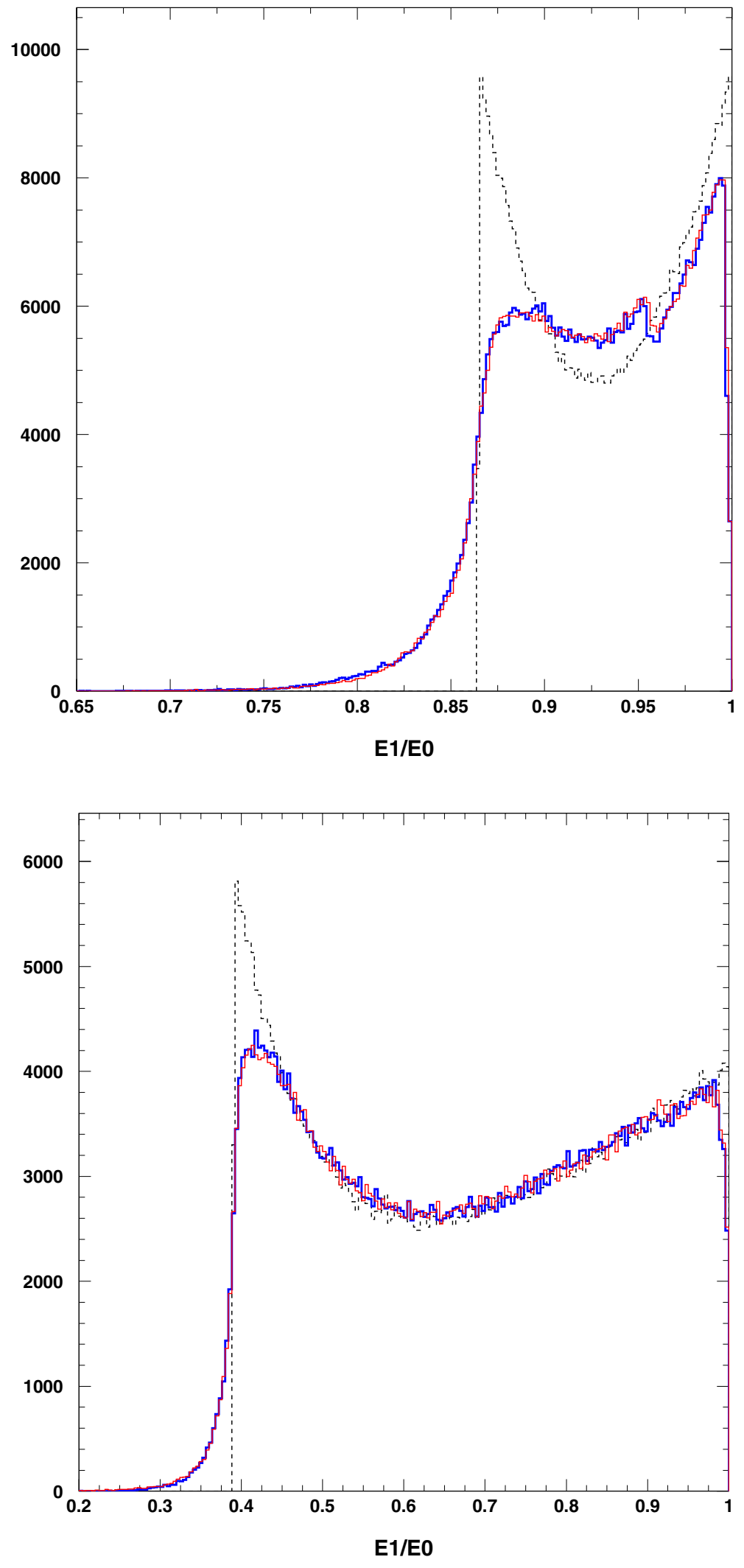

Figure 7: Final state generators of the Library, Penelope and Standard flavors of Compton scattering, for two different elements and incident photon energies: Si, 40 keV (top) and Cs, 400 keV (bottom), respectively for Library based (blue), Penelope-like (red) and Standard (black) final state modelling.

Table 1: $\chi^2$ test for the three photon interaction models in Geant4 with respect to the NIST reference data.

| Model | p-value |
|---|---|
| Library-based | 0.982 |
| Penelope-like | <0.001 *(0.993 excluding the datum at 1 keV)* |
| Standard | 0.189 |

## IV. Computational Performance

Preliminary results of the electromagnetic physics pilot project indicate a performance improvement associated with the policy-based design.

The results for different materials are shown in table 2 for the Penelope-like flavor of Compton scattering; similarly, the Library-based implementation according to the policy-based design shows 28% gain with respect to the equivalent implementation in Geant4 low energy electromagnetic package. The simulation is based on 40 keV photons, $10^6$ events, Intel Core2 Duo Processor E6420, 2.13 GZ, 4 GB RAM.

Table 2: Percentage of gain in performance.

| Element | Policy-based design | Geant4 9.1 | Gain |
|---|---|---|---|
| C | 4.15 | 6.08 | 32% |
| Si | 6.23 | 8.37 | 26% |
| Cu | 7.64 | 10.78 | 29% |
| W | 14.06 | 19.18 | 27% |

## CONCLUSION

This project addresses various features relevant to experimental simulation applications, like the relative physics characteristics and computational speed of Geant4 models of photon interactions.

The prototype design technique adopted in the pilot project of photon interaction design looks promising for application to a large, complex, computationally intensive physics simulation domain. Meaningful gain is achieved in transparency, agility, ease of verification and validation, and software maintenance. Significant performance improvements have been observed at a very early stage of the project, still leaving room for further ones.

The documentation of the full set of results will be made available to the experimental community as soon as the Verification & Validation process is completed.